\documentclass[reprint, aps]{revtex4-1}
\usepackage{amsmath}
\usepackage{graphicx}
\usepackage{dcolumn}
\usepackage{bm}
\usepackage{float}
\usepackage{appendix}
\usepackage{lettrine}
\usepackage{graphics}
\usepackage{amssymb}
\usepackage{color}
 \usepackage {marvosym}
 \usepackage{ifsym}
\definecolor{myred}{rgb}{0.75,0,0}
 \usepackage{natbib}
\usepackage{bbm}
\usepackage{ulem}
\usepackage{color}
\definecolor{mygreen}{rgb}{0,0.5,0}
\definecolor{mygrey}{rgb}{0.5,0.5,0.5}
\definecolor{myred}{rgb}{0.75,0,0}
\definecolor{myblue}{rgb}{0,0,0.75}
\definecolor{mymagenta}{cmyk}{0,1,0,0.12}
\definecolor{mycyan}{cmyk}{1,0,0,0.12}
\definecolor{myorange}{rgb}{1,0.5,0}
\definecolor{myviolet}{rgb}{0.5,0.0,0.75}
\definecolor{mybrown}{rgb}{0.75,0.5,.5} 
\usepackage{amsthm,amsmath,amssymb}
\usepackage{mathrsfs}

\renewcommand{\and}{a_n^\dagger}

\begin{document}
\newcommand{\thetitle}{Quantum Twin Interferometers}
	\title{\thetitle}
 \author{Wei Du $^{1,2} \dagger$}
\author{Shuhe Wu $^{1,2} \dagger$}
\author{Dong Zhang $^{1,2}$}
\author{Jun Chen $^{1,2}$}
\author{Yiquan Yang $^{1,2}$}
\author{Peiyu Yang $^{1,2}$}
\author{Jinxian Guo $^{1,2}$}
\author{Guzhi Bao $^{1,2\,*}$ }
\author{Weiping Zhang $^{1,2,3,4\,*}$}
\affiliation{
$^{1}$ School of Physics and Astronomy, and Tsung-Dao Lee institute, Shanghai Jiao Tong University, Shanghai 200240, China.\\
$^{2}$ Shanghai Research Center for Quantum Sciences, Shanghai 201315, China.\\
$^{3}$ Shanghai Branch, Hefei National Laboratory, Shanghai 201315, China.\\
$^{4}$ Collaborative Innovation Center of Extreme Optics, Shanxi University, Taiyuan, Shanxi 030006, China.\\
$\dagger$ These authors contributed equally to this work.\\
$^{*}$Corresponding author. Email: guzhibao@sjtu.edu.cn; wpz@sjtu.edu.cn.}
\date{\today}
\newcommand{\Suppl}{Supplementary information}
\begin{abstract}
{ Quantum-correlated interferometer is a newly emerging tool in quantum technology that offers classical-limit-breaking phase sensitivity. But to date, there exists a configurational bottleneck for its practicability due to the low phase-sensitive photon numbers limited by the current detection strategies. Here we establish an innovative development termed as ``$quantum$ $twin$ $interferometer$'' with dual pairs of entangled twin beams arranged in the parallel configuration, allowing fully exploits the quantum resource through the new configuration of entangled detection. We observe the distributed phase sensing with 3\,dB quantum noise reduction in phase-sensing power at the level of milliwatts, which advances the record of signal-to-noise ratio so far achieved in photon-correlated interferometers by three orders of magnitude. The developed techniques in this work can be used to revolutionize a diversity of quantum devices requiring phase measurement.}  
\end{abstract}
\pacs{Valid PACS appear here}
\maketitle
\textbf{Introduction}
\\
Lasers are ideal carriers of information due to their advantages of coherence, monochromaticity, and ability to maintain a stable phase relation over long distances \cite{RevModPhys.71.S471}. Various physical quantities can be converted into an optical phase shift, making the sensitive estimation of phase variation as an essential goal in precision measurement \cite{RevModPhys.90.035005}. Over the past several decades, interferometric techniques have developed well into indispensable tools for phase measurement and been extensively applied in numerous fields, including navigation \cite{gyroscope1,LaiYuHung}, field sensing \cite{RN1,PhysRevApplied.20.064028}, and gravitational wave observatory \cite{RevModPhys.52.285,PhysRevLett.116.061102}, etc. 
For a linear interferometer with coherent injection, the vacuum fluctuations in input states make the phase sensitivity bounded by the standard quantum limit (SQL), scaling as $ 1/\sqrt{N}$, where $N$ is the mean photon number for phase sensing \cite{PhysRevD.23.1693}.

\begin{figure}[t]
	\centering\includegraphics[width=8.6cm]{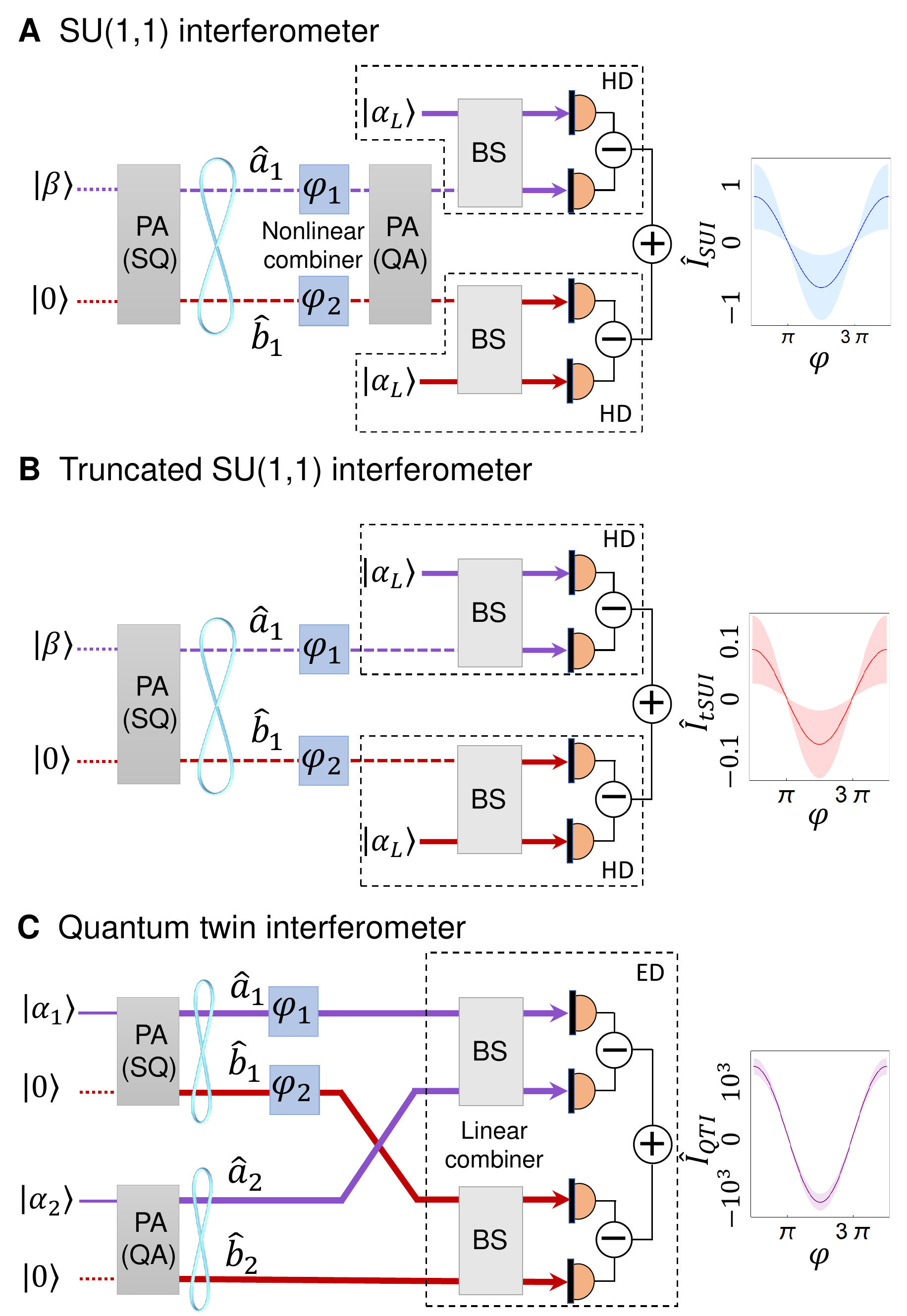} 	\caption{\textbf{The protocols and schematic diagram of Quantum Twin Interferometer. } \textbf{(A)}, SU(1,1) interferometer with $I_{ps}=|\cosh(2s)\beta|^{2} $ which is limited by nonlinear combiner. \textbf{(B)}, Truncated SU(1,1) interferometer with $I_{ps}=|\cosh(2s)\beta|^{2}\ll|\alpha_{L}|^{2}$ which is limited by the configuration of HD to avoid the fluctuations from the classical reference beams. Here $|\beta|^{2}$ is the seed power for SU(1,1) and tSUI, $|\alpha_{L}|^{2}$ is the power of reference beam. \textbf{(C)}, Quantum twin interferometer with $ I_{ps}=\left[\cosh(s)\alpha_{1}\right]^{2}+\left[\sinh(s)\alpha_{2}\right]^{2}=|\cosh(2s)\alpha|^{2}/2$ when $|\alpha_{1}|=|\alpha_{2}|=|\alpha|/\sqrt{2}$, $s_{1}=s_{2}=s$, which eliminate the influence of classical fluctuations originating from the reference beams in the tSUI benefit by the effective utilization of entangled reference beams in entangled detection (ED). Here, $s$ is the squeezing parameter for all PAs. The right plot shows the output photocurrent for each strategy. Shading indicates the noise fluctuation.  $\varphi=\varphi_{1}/2+\varphi_{2}/2$ is the total variation of phase. }
	\label{fig_1}
\end{figure}
The development of nonlinear optics enables the generation of nonclassical states with quantum correlation and manipulation of the quantum statistics of photons \cite{DFWAlls,PhysRevLett.57.2520,PhysRevLett.55.2409,PhysRevLett.117.110801}, leading to the advanced interferometry for detecting signals immersed in vacuum fluctuation \cite{PhysRevLett.59.278,PhysRevLett.71.1355,RN11}. The earliest experiment to surpass the SQL is to squeeze the vacuum fluctuation by replacing the unused input port of a linear interferometer with a squeezed state \cite{PhysRevLett.59.278}. This strategy has been experimentally demonstrated and recently implemented in the advanced Laser Interferometer Gravitational-Wave Observatory (LIGO) \cite{LIGO1,LIGO2}. Unfortunately, quantum correlation is so fragile facing realistic environments that only a limited quantum enhancement can be achieved in practice. 
As seen in experiments of linear interferometers with squeezed state injection, the squeezing factor is considerably reduced due to the losses from optical elements, optical path, misalignment, and mode mismatch \cite{PhysRevLett.59.278,PhysRevLett.123.231107}.

\begin{figure*}[t]
\centering
\includegraphics[width=18cm]{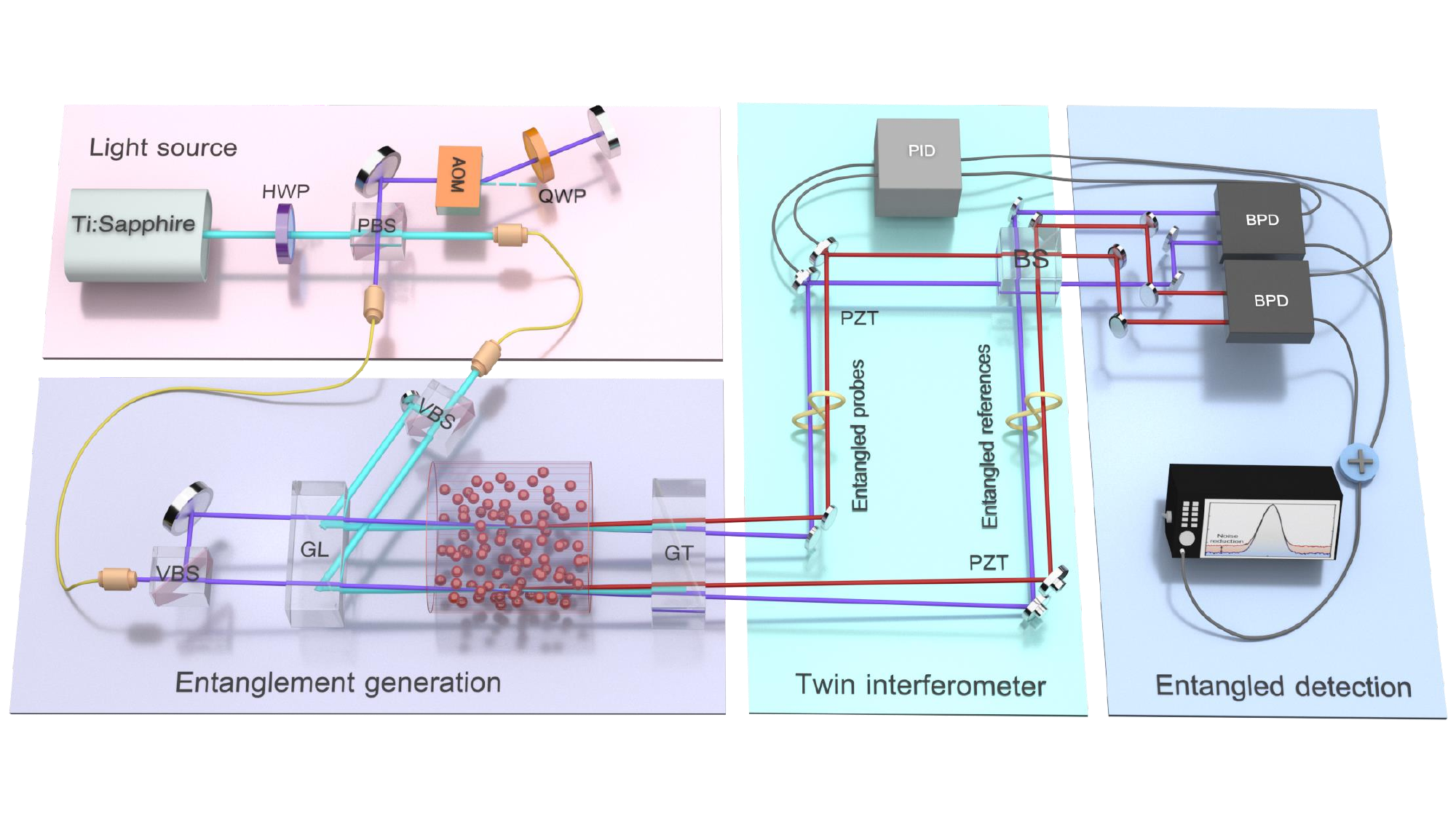} 
\caption{\textbf{Schematic diagram of the experiment for twin interferometer.} PBS: polarization beam splitter; VBS: variable beam splitter. BS: beam splitter; HWP: half-wave plate; QWP: quarter-wave plate;
PZT: piezoelectric transducer; AOM: Acousto-optic modulators; GL: Glan-Laser polarizer; GT: Glan–Thompson polarizer. PID: proportion integration differentiation; BPD: balanced photo-detector;} 
\label{fig:setup}
\end{figure*}

An alternative approach is to integrate the squeezing and the beam splitting into a single nonlinear optical process, directly generating and utilizing entangled light beams for phase sensing beyond the SQL even in the presence of losses.
This can be achieved through a nonlinear interferometer called SU(1,1) interferometer (SUI) \cite{PhysRevA.33.4033,plick2010coherent,SUI1}, consisting of two cascaded parametric amplification (PA) processes. The initial PA serves as the squeezer (SQ) generating entangled beams and another PA, actually a quantum amplifier (QA), here functions as a nonlinear combiner for these beams, as illustrated in Fig.~\ref{fig_1}A. 
The SUI has been successfully demonstrated in various systems, e.g. a full-optical one \cite{SUI1, PhysRevLett.119.223604,Liu:19,PhysRevLett.128.033601}, an atom-light hybrid one \cite{PhysRevLett.115.043602} and an atomic one \cite{gross2010nonlinear,PhysRevLett.117.013001,liu2022nonlinear}, evidently exhibiting the tolerance to losses\cite{SUI1,PhysRevLett.119.223604,Liu:19}. However, a notable drawback of the SUI lies in its nonlinear combiner, which induces excess losses from mode mismatch and amplifies uncorrelated noises \cite{PhysRevA.86.023844}, while also imposing a restriction of optical power on the phase-sensing light \cite{pooser2009quantum}. This limitation hinders the full exploitation of quantum resources.

Recently, a modified version of SUI is proposed and demonstrated that replaces the nonlinear combiner with dual homodyne detection (HD), as shown in Fig.~\ref{fig_1}B \cite{Anderson:17}. This strategy, termed as truncated SUI (tSUI) offers a facilitated configuration compared to the full SUI, eliminating drawbacks associated with the nonlinear combiner in SUI. But the story is far from over. A new barrier from HD is brought to tSUI. 
In the HD, a reference beam with a much stronger power than that of the probe beam is required to amplify the probe beam, avoiding the field fluctuation of the reference beam. 
In this sense, the HD actually acts as an imbalanced Mach-Zehnder interferometer (MZI) \cite{PhysRevA.95.063843} between the reference beam and the probe one with a low fringe visibility due to imbalanced interference.
This results in a low utilization efficiency of photon resources in tSUI and poor capability to achieve high photon numbers for phase sensing (see Supplementary materials Text\,I).
Consequently, despite its successful implementation in various applications, the sensitivity of tSUI remains far from satisfactory, as the probe beam's power has not been fully exploited to its upper limit in these cases \cite{taylor2016quantum,PhysRevLett.124.230504,QEmicroscopy,NPRS,NMfluore}. 

To fully exploit photon resources in such interferometry for phase sensing with high photon number, a new detection configuration is desired to replace the HD.
In this paper, we present an innovative configuration, ``quantum twin interferometer'' (QTI), as shown in Fig.~\ref{fig_1}C, where a second interferometer (colored in purple) is cloned from an original one (colored in red) through the utilization of two PAs. Diverging from the cascaded PAs found in the SUI, the QTI employs PAs arranged in parallel. By substituting the nonlinear combiner in the SUI with a linear one, the QTI effectively circumvents the drawbacks associated with the QA, akin to the tSUI. 

More importantly, the QTI, employing dual pairs of entangled twin beams generated from the PAs, actually acts as a pair of correlated interferometers. In contrast to the HD configuration in the tSUI \cite{Zhou:23},  the probe and reference with equal power allow a balance interference with perfect fringe visibility. 
Meanwhile, the entanglement between reference beams leads to the entangled detection (ED) in the interferometry, suppressing the field fluctuations in reference beams (see Supplementary materials Text II).

Breaking through the constraints of both SUI and tSUI, this comprehensive design enables the QTI to accommodate a significant amount of phase-sensing power within the interferometer arms while suppressing noises. Experimental results reveal that the QTI reduces quantum noise by 3\,dB, while increasing the phase-sensing power $I_{ps}$ by three orders of magnitude compared to previously reported photon-correlated interferometers. 
Our work of QTI revolutionizes the optical interferometry with a new method of entangled detection, pushing the sensitivity to regimes that previous photon-correlated interferometers struggled to achieve \cite{SUI1,PhysRevLett.117.013001,PhysRevLett.119.223604,Liu:18,Anderson:17,PhysRevLett.124.173602,PhysRevLett.124.230504}, thereby opening up new prospects for practical applications.
\\
\\
\textbf{Results}
\\
In our system, the PAs are achieved by the four-wave-mixing (FWM) process for generating the twin beams, whose interaction Hamiltonian is
\begin{eqnarray}\label{Ham}
\hat{H}_{n}=i\hbar\xi\hat{a}_{n}^{\dagger}\hat{b}_{n}^{\dagger}+h.c.,
\end{eqnarray}
where $n\in(1,2)$ represents the photons generated from SQ or QA. $ \xi $ is the strength of interaction which depends on pump power, one-photon detuning, two-photon detuning, etc. 

As illustrated in Fig.~\ref{fig_1}C, a pair of PAs with seed powers $ |\alpha_{1}|^{2} $ and $ |\alpha_{2}|^{2} $ generate probe and reference beams. In general, the seed powers and gains for two PAs are not necessary to be the same. For convince, we name this general case as pair quantum correlated interferometry (PQCI). The resulting output observable is
\begin{eqnarray}\label{tc}
\hat{I}_{PQCI}=\hat{a}_{1}^{\dagger}\hat{a}_{2}e^{i\varphi_{1}}+\hat{a}_{2}^{\dagger}\hat{a}_{1}e^{-i\varphi_{1}}+\hat{b}_{1}^{\dagger}\hat{b}_{2}e^{i\varphi_{2}}+\hat{b}_{2}^{\dagger}\hat{b}_{1}e^{-i\varphi_{2}},\nonumber\\
\end{eqnarray}
where $\varphi_{1}$ and $\varphi_{2}$ are the relative phase of the original interferometer and cloned interferometer.

For optical interferometers, the phase sensitivity can be determined by monitoring the changes in light intensity. As illustrated in Fig~\ref{fig_1}, the results depend on the phases $ \varphi_{1}=\varphi_{10}+\delta\varphi_{1} $ and $ \varphi_{2}=\varphi_{20}+\delta\varphi_{2} $. Here, $ \varphi_{10} $ and $\varphi_{20}$ represent the reference points for each interferometer, while $ \delta\varphi_{1} $ and $  \delta\varphi_{2}$ are the small signals to be measured. The assessment of the system's ability to detect the unknown phases is done by computing the signal-to-noise ratio (SNR) of the output signal, defined as
\begin{eqnarray}\label{dsnr}
\zeta=\frac{(\left\langle I\right\rangle -\left\langle I\right\rangle_{0} )^{2}}{\left\langle \delta^{2}I\right\rangle_{0}},
\end{eqnarray}
where $ \left\langle I\right\rangle _{0} $ is the expectation of output taken with $ \varphi_{1}=\varphi_{10} $ and $ \varphi_{2}=\varphi_{20} $, indicating an undisturbed interferometer. $ \left\langle I\right\rangle  $  signifies the expectation with $ \varphi_{1}=\varphi_{10}+\delta\varphi_{1} $ and $ \varphi_{2}=\varphi_{20}+\delta\varphi_{2} $, in the presence of weak signals causing slight phase excursion in the interferometer. $ \left\langle \delta^{2}I\right\rangle_{0}  $ represents the variance when the interferometer is set at $ \varphi_{1} =\varphi_{10}$ and $ \varphi_{2} =\varphi_{20}$.

Then we can acquire the SNR of PQCI from Eq.~\ref{dsnr}
\begin{eqnarray}\label{tc}
\zeta_{PQCI}=\frac{4\mathcal{R}(1-\mathcal{R})\lbrace\left[ \cosh^{2}(s)\delta\varphi_{1}+\sinh^{2}(s)\delta\varphi_{2}\right] \alpha^{2}\rbrace ^{2}}{\alpha^{2}},\nonumber\\
\end{eqnarray}
where $s$ is the squeezing parameter for both PAs. $\mathcal{R}=|\alpha_{1}|^{2}/(|\alpha_{1}|^{2}+|\alpha_{2}|^{2})$ represents the power ratio of one seed beam to the total seed beams in PQCI. In tSUI, the twin beams are detected with classical local references. Therefore, quantum-enhanced sensing is achievable only when $\mathcal{R}\rightarrow0$ with the HD configuration, which ignores the fluctuations of the reference beams (see Supplementary materials Text\,I), leading to significantly weak signal strength. In PQCI, both the signal and reference beams are correlated, leading to the output noise independent of the proportion of the seed injection $\mathcal{R}$. To achieve high absolute sensitivity, it is crucial to effectively utilize the photons within the interferometer. 
When $\mathcal{R}=1/2$, the PQCI ends up as the QTI, achieving maximum signal and squeezing simultaneously, which leads to the optimal SNR $\zeta_{QTI}=\left[\cosh^{2}(s)\delta\varphi_{1}+\sinh^{2}(s)\delta\varphi_{2}\right]^{2}\alpha^{2}$.
This advancement removes the previous constraints associated with the nonlinear combiner in SUI and the HD in tSUI, achieving an absolute sensitivity that surpasses the previously reported levels in SUI and tSUI.

\begin{figure*}[t]
\centering
\includegraphics[width=18cm]{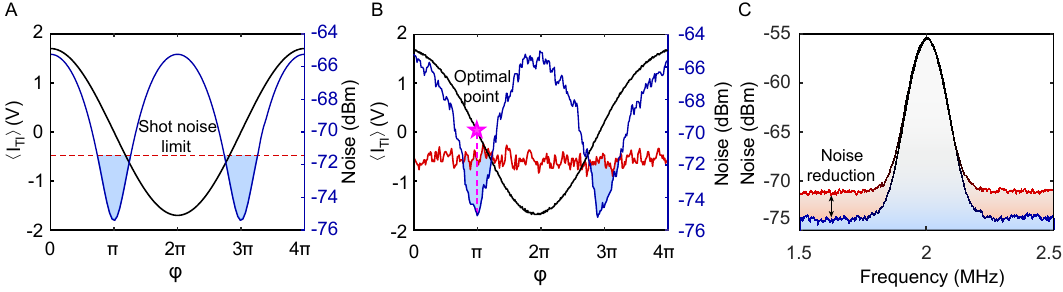} 
\caption{\textbf{Comparison of spectrum analysis between quantum twin interferometer (QTI) and classical Mach Zehnder interferometer (MZI) performance in phase measurement with the same $ I_{ps}$.} \textbf{(A)}, theoretical noise power spectra. \textbf{(B)}, experimental noise power spectra. Black: signal of QTI with scanned global phases ($ \varphi=\varphi_{1}/2+\varphi_{2}/2$); Blue: noise of QTI with scanned global phase; Red: MZI locked at minimal noise, indicating shot noise limit. \textbf{(C)}, SNR comparison between QTI and MZI with same phase signal at 2 MHz. Black $\rightarrow$ Red: MZI; Black $\rightarrow$ Blue: QTI. Traces are recorded with a 100 kHz resolution bandwidth (RBW) and 300 Hz video bandwidth (VBW). Traces in \textbf{(B)} are averaged three times.} 
\label{fig:3}
\end{figure*}


The experimental arrangement for QTI is shown in Fig.~\ref{fig:setup}. Two sets of twin beams are generated by two non-degenerate FWM processes \cite{PhysRevA.78.043816}. In this experiment, the pump beam is supplied by a Ti: sapphire laser, whose frequency is locked at blue-shifted by approximately $ \Delta=1 $\,GHz above the transition line of the D1 line of $ ^{85} $Rb 5\,S$_{1/2} $ → 5\,P$_{1/2} $, 795\,nm (see Supplementary Material Text\,IV). The seed light is red-shifted by 3.38\,GHz from the pump beams through the double pass configuration of a 1.5\,GHz acoustic-optic modulator (AOM). The weak seed beam with a waist of 250\,$\mu$m intersects a strong pump beam with a waist of 500\,$\mu$m at an angle of 0.3$ ^{\circ}$ in a 12\,mm long $ ^{85} $Rb vapor cell maintained at a temperature of 120\,$ ^{\circ}$C. Two pairs of correlated photons are generated after the FWM processes. The correlated photons are normally referred to as `signal' and `idler' with approximately a 6\,GHz frequency difference, denoted by modes $ \hat{a}_{n} $ and $ \hat{b}_{n} $ with $n\in(1,2)$ representing photons from the first or second FWM processes as outlined in Eq.~\ref{Ham}. The correlated photon with signal and idler frequency are separately combined at BSs, and the output light is then sent to differential detectors. The DC parts of the subtracted currents are injected into Proportion-Integration-Differentiation (PID) controllers and feedback to the arms of QTI for locking the phase \cite{DuOL}. All the phase shifts in this paper are achieved by piezoelectric transducers (PZTs). The AC parts of the differential currents are summed and sent to the spectrum analyzer (SA). The seed power can be adjusted by rotating the half-wave plate before the Glan-Laser polarizer in front of the cells, making it easy to switch between QTI and another strategy of PQCI by setting equal or highly unbalanced seed power for the two FWM processes. It is also flexible to remove the cells, resulting in a switch to the classical MZI achieved by the interference of two seed lights.

The enhanced performance of QTI compared to classical MZI with the same $I_{ps}$ is illustrated in Fig.~\ref{fig:3}\,A and B represent the theoretical and experimental results of their noise power spectra performance respectively. The red traces indicate the noise level of MZI with the same $I_{ps}$ compared to QTI. The black and blue traces show the signal and noise of QTI respectively when scanning the phase within the interferometer. This reveals a minimal noise level, corresponding to approximately 3.5\,dB suppression in noise compared to MZI at the same $I_{ps}$. Fig.~\ref{fig:3}\,C illustrates the SNR comparison between QTI and MZI for measuring the phase of $\delta\varphi_{1}=\delta\varphi_{2}$ modulated at 2\,MHz. MZI and QTI yield SNR values of 15.5\,dB and 18.5\,dB respectively, showing a 3\,dB improvement. The results of QTI above were obtained with the gain of PAs $G_{q} = \cosh^2(s) = 3$, where both PAs have the same seed power and the $ I_{ps}=\cosh(2s)\alpha^{2}=400\,\mu$W.

\begin{figure*}[t]
\centering
\includegraphics[width=18cm]{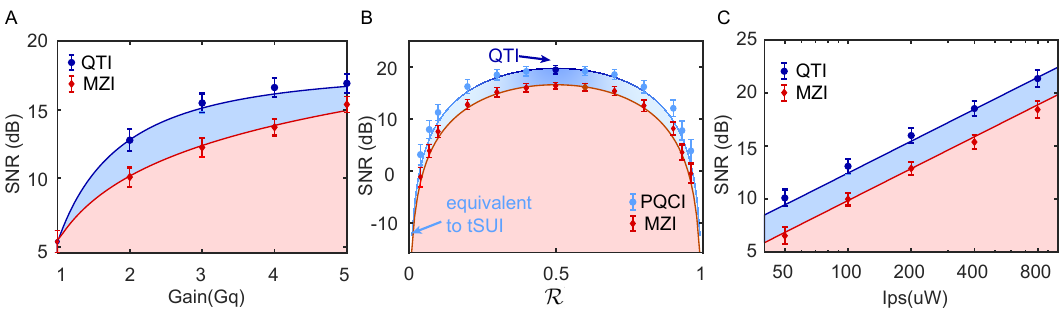} 
\caption{\textbf{Optimizing the SNR in phase measurement and comparing the performance between QTI and MZI.} Blue: QTI; Red: MZI. \textbf{(A)}, SNR versus gain G$_{q}$. \textbf{(B)}, SNR versus the proportion of probe light within the total light intensity inside the interferometer ($\mathcal{R}$). When $\mathcal{R}=0.5$, PQCI becomes QTI. When $\mathcal{R}\rightarrow\,0$, PQCI is equivalent to tSUI. \textbf{(C)}, SNR versus the intensity of phase-sensing light (Ips).} 
\label{fig:4}
\end{figure*}

To determine the optimal operating conditions, the evolution as a function of various experimental parameters is depicted in Fig.~\ref{fig:4}. Here, the SNR of QTI and MZI are both run at the same $I_{ps}$. In Fig.~\ref{fig:4}A, measurements are taken with $\mathcal{R}=1/2$ and $I_{ps}=400\mu$W. SNR rises with increasing $G_{q}$, reaching saturation around $ G_{q}\rightarrow5 $. The maximal quantum enhancement occurs when $G_{q} = \cosh^2(s) = 3$, decreasing slightly with each subsequent increase in $ G_{q} $. This phenomenon can be described by the imperfect mode matching between spatial multi-mode twin beams introducing uncorrelated thermal noise \cite{Gupta:20}. The SNR versus $\mathcal{R}$ is depicted in Fig.~\ref{fig:4}B. It is easy to observe that quantum enhancement can be maintained with all $\mathcal{R}$. Nevertheless, when the seed power for the two PAs is highly imbalanced, the majority of the amplified photons go toward the reference beams. In this case, the quantum noise of the reference can be ignored, resulting in PQCI being equivalent to tSUI, which exhibits a notably low SNR in such a strategy.
Fig.~\ref{fig:4}C illustrates the relationship between SNR and $I_{ps}$, which grows with the increase in seed injection power. The results reveal a proportional growth in SNR with $I_{ps}$, and the 3\,dB quantum enhancement persists as $I_{ps}$ approach the order of milliwatts, a magnitude significantly greater than the sub $ \mu $W levels observed in previous tSUIs \cite{Anderson:17,PhysRevLett.124.173602}. A detailed theory (refer to the supplementary Material\,III) that takes into account the loss from optical path and modes mismatching fits the experimental data well, as indicated by the solid lines in Fig.~\ref{fig:4}.
\\
\\
\textbf{Discussion}
\\
In summary, we employ two pairs of twin beams to construct a pair of correlated interferometers as QTI. We notice that the construction of QTI is similar to that of tSUI, both of which have intrinsic mechanisms that build correlations between a pair of interferometers to achieve quantum-enhanced precision measurements. Both tSUI and QTI feature a facilitated setup compared to the previous quantum interferometer \cite{SUI1,PhysRevLett.124.173602} by combining interferometry and signal readout processes, reducing losses such as those from linear and nonlinear mixing. Compared to the weak signal of tSUI due to the restrictions of classical reference beams, QTI allows for correlating a pair of interferometers with balanced intensity between the arms, resulting in significantly increased signal strength while preserving all the benefits of the tSUI.

The proposed device exploits positive phase signal correlation and inverse noise correlation at its operating point, allowing for distributed sensing and correlated noise cancellation by summing the outputs of the correlated interferometers. We observed a 3\,dB quantum enhancement in $I_{ps}$, which is three orders of magnitude higher than previously reported for tSUI. In principle, the sensitivity could approach the Heisenberg scaling by reducing the number of seed photons (see Supplementary Material Text\,II. The robustness of the quantum enhancement across a wide range of $I_{ps}$ offers flexibility for different types of practical applications, such as bio-sensing \cite{QEmicroscopy,SERS} and ultrasensitive measurements of force \cite{PhysRevLett.124.230504,QENMZ}. In contrast to previously reported photon-correlated interferometers \cite{SUI1,Anderson:17,PhysRevLett.124.173602}, this method does not require additional HD, thus eliminating the need for extra local oscillator fields and the associated mode-matching issues. This makes the cloning method more practical and readily extendable to other types of optical systems.
\\
\\
\textbf{REFERENCES}
\bibliographystyle{apsrev4-1no-url}
\bibliography{main1}
\textbf{ACKNOWLEDGMENTS}
\\
\textbf{Funding:} This work was supported by the Innovation Program for Quantum Science and Technology (2021ZD0303200), the National Natural Science Foundation of China (Grant No. 12234014, 12374331, 12204304, 11904227, 12404416 and 11654005), the Shanghai Municipal Science and Technology Major Project (2019SHZDZX01), the Natural Science Foundation of Shanghai (24ZR1437900), the Fellowship of China Postdoctoral Science Foundation (Grant No. 2020TQ0193, 2021M702146, 2021M702150, 2021M702147, 2022T150413, GZB20230424), and the National Key Research and Development Program of China under Grant number 2016YFA0302001. W.Z. also acknowledges additional support from the Shanghai talent program.
\textbf{Author contributions:} W.Z. supervised the whole project. W.D., G.Z.B., and W.Z. conceived the research. G.Z.B, W.D., S.H.W, and W.Z. designed the experiments. W.D., S.H.W., D.Z., J.C., Y.Q.Y.  .Y.Y., J.X.G. and G.Z.B. performed the experiment. G.Z.B., W.D., S.H.W, and W.Z. contributed to the theoretical study. G.Z.B., W.D., and S.H.W. analyzed the data. G.Z.B., W.D., S.H.W., and J.C. draw the diagrams. G.Z.B., W.D., S.H.W, and W.Z. wrote the paper. All authors contributed to the discussion and review of the manuscript. 
\textbf{Competing interests:} The authors declare no competing financial interests. \textbf{Data andmaterials availability:} All data needed to evaluate the conclusions in the paper are present in the paper or the supplementary materials. \textbf{License information:} Copyright © 2024 the authors, some rights reserved; exclusivelicensee American Association for the Advancement of Science.
\\
\\
\\
\\
\textbf{SUPPLEMENTARY MATERIALS}
\\
Supplementary Text\\
\begin{widetext}
	\noindent
 \section{The SNR of truncated SU(1,1) interferometer}
The configuration of truncated SU(1,1) interferometer (tSUI) [26], is shown in Fig.1B. Coherent light are employed as reference beams to enhance the photon numbers within the interferometer. The intensity of the light field at the output can be expressed as
\begin{equation}\label{tic}
\hat{I}_{tSUI}=\hat{a}_{1}^{\dagger}\hat{a}_{lo}e^{i\varphi_{1}}\!+\!\hat{a}_{lo}^{\dagger}\hat{a}_{1}e^{-i\varphi_{1}}\!+\!\hat{b}_{1}^{\dagger}\hat{b}_{lo}e^{i\varphi_{2}}\!+\!\hat{b}_{lo}^{\dagger}\hat{b}_{1}e^{-i\varphi_{2}}\nonumber\\\tag{S1}
\end{equation}
Where $ \hat{a}_{lo} $ and $ \hat{b}_{lo} $ denote the modes for the reference light. The optimal sensitivity is achieved when $ \varphi_{10}=\pi/2 $ and $ \varphi_{20}=\pi/2 $. The corresponding SNR is
\begin{equation}\label{snrtbi}
\zeta_{tSUI}=\frac{2|\alpha_{L}|^{2}|\beta|^{2}\left[ \cosh(s)\delta\varphi_{1}+\sinh(s)\delta\varphi_{2}\right]^{2}}{|\alpha_{L}|^2e^{-2s}+(|\beta|^{2}+1)\cosh{2s}-1}\tag{S2}
\end{equation}
Here $s$ is the squeezing parameter of the PA. 
We notice that the fluctuations of the classical reference beams will contribute to the noise when the power of the reference beam $|\alpha_{L}|^2$ is comparable to the power of probe beam $\cosh(2s)|\beta|^2$, resulting in classical features gradually dominating the noise part. The uncorrelated noise can be eliminated, allowing for the effective use of quantum squeezed noise while $|\alpha_{L}|\gg|\beta|$, leading to $\left\langle\delta^{2} \hat{I}_{tSUI}\right\rangle_{0}=|\alpha_{L}|^{2}e^{-2s}+(|\beta|^{2}+1)\cosh(2s)-1 \approx|\alpha_{L}|^{2}e^{-2s}$. In this scenario, the output photocurrent can be approximated as $I_{tSUI}\propto\sqrt{|\alpha_{L}|/2}\left[\hat{X}_{\hat{a}_{1}}(\varphi_{1})+\hat{X}_{\hat{b}_{1}}(\varphi_{2})\right]$. This represents a dual homodyne configuration, where the majority of photons serve as the reference beams, with only a small fraction dedicated to measurement. Considering the requirement for highly imbalanced power of probe and reference beams $|\alpha_{L}|\gg|\beta|$, this approach functions as a combination of two extremely unbalanced interferometers, resulting in a very weak signal. Currently, obtaining a high SNR poses a challenge due to the ineffective utilization of the injected coherent light.
\section{Theoretical calculation of quantum twin interferometer}

The quantum optical performance of the Quantum Twin Interferometer (QTI) can be analyzed by considering cascaded linear input-output relations, i.e., through a series of linear transformations on field operators.
\begin{equation}\label{Input-output relation}
\hat{a}_{s}=\sqrt{\mathcal{R}}\hat{a}_{0}+\sqrt{1-\mathcal{R}}\hat{b}_{0} \tag{S3}
\end{equation}
\begin{equation}\label{Input-output relation}
	\hat{b}_{s}=\sqrt{\mathcal{R}}\hat{b}_{0}-\sqrt{1-\mathcal{R}}\hat{a}_{0}\tag{S4}
 \end{equation}
\begin{equation}\label{Input-output relation}
\hat{c}=\cosh(s_{1})\hat{a}_{s}+\sinh(s_{1})\hat{a}_{i}^{\dagger}\tag{S5}
 \end{equation}
\begin{equation}\label{Input-output relation}	\hat{d}=\cosh(s_{1})\hat{b}_{i}+\sinh(s_{1})\hat{b}_{s}^{\dagger}\tag{S6}
 \end{equation}
\begin{equation}\label{Input-output relation}	\hat{e}=\cosh(s_{2})\hat{b}_{s}+\sinh(s_{2})\hat{b}_{i}^{\dagger}\tag{S7}
 \end{equation}
\begin{equation}\label{Input-output relation}	\hat{f}=\cosh(s_{2})\hat{a}_{i}+\sinh(s_{2})\hat{a}_{s}^{\dagger}\tag{S8}
 \end{equation}
 \begin{equation}\label{Input-output relation}
	\hat{g}=\frac{1}{\sqrt{2}}(e^{i\varphi_{1}}\hat{c}+\hat{e})\tag{S9}
 \end{equation}
 \begin{equation}\label{Input-output relation}
	\hat{h}=\frac{1}{\sqrt{2}}(e^{i\varphi_{1}}\hat{c}-\hat{e})\tag{S10}
 \end{equation}
 \begin{equation}\label{Input-output relation}
	\hat{i}=\frac{1}{\sqrt{2}}(e^{i\varphi_{2}}\hat{f}+\hat{d})\tag{S11}
 \end{equation}
 \begin{equation}\label{Input-output relation}
	\hat{j}=\frac{1}{\sqrt{2}}(e^{i\varphi_{2}}\hat{f}-\hat{d})\tag{S12}
\end{equation}	
where $s_{1,2}$ symbolizes the squeezing parameter. Here, $\mathcal{R}=|\alpha_{1}|^{2}/(|\alpha_{1}|^{2}+|\alpha_{2}|^{2})$ 
with $|\alpha_{1}|^{2}+|\alpha_{2}|^{2}=|\alpha|^{2}$ represents the power ratio of one seed beam to the total seed beams in pair quantum correlated interferometry (PQCI). Ultimately, the output modes are simplified to
\begin{equation}\label{general output}
	\hat{g}=\frac{1}{\sqrt{2}}\left\lbrace\left[\left(\cosh(s_{1})\sqrt{\mathcal{R}}e^{i\varphi_{1}} -\cosh(s_{2})\sqrt{1-\mathcal{R}}\right)\hat{a}_{0}+\left(\cosh(s_{1})\sqrt{1-\mathcal{R}}e^{i\varphi_{1}}+\cosh(s_{2})\sqrt{\mathcal{R}}\right)\hat{b}_{0}\right]+\sinh(s_{1})e^{i\varphi_{1}}\hat{a}_{i}^{\dagger}+\sinh(s_{2}){b}_{i}^{\dagger}\right\rbrace \nonumber\tag{S13}
 \end{equation}	
 \begin{equation}\label{general output}
	\hat{h}=\frac{1}{\sqrt{2}}\left\lbrace\left[\left(\cosh(s_{1})\sqrt{\mathcal{R}}e^{i\varphi_{1}}+\cosh(s_{2})\sqrt{1-\mathcal{R}}\right)\hat{a}_{0}+\left(\cosh(s_{1})\sqrt{1-\mathcal{R}}e^{i\varphi_{1}}-\cosh(s_{2})\sqrt{\mathcal{R}}\right)\hat{b}_{0}\right]+\sinh(s_{1})e^{i\varphi_{1}}\hat{a}_{i}^{\dagger}-\sinh(s_{2}){b}_{i}^{\dagger}\right\rbrace\nonumber\tag{S14}
 \end{equation}	
 \begin{equation}\label{general output}
	\hat{i}=\frac{1}{\sqrt{2}}\left\lbrace \left[\left(\sinh(s_{1})\sqrt{\mathcal{R}}e^{i\varphi_{2}}-\sinh(s_{2})\sqrt{1-\mathcal{R}}\right)\hat{a}_{0}^{\dagger}+\left(\sinh(s_{1})\sqrt{1-\mathcal{R}}e^{i\varphi_{\varphi_{2}}}+\sinh(s_{2})\sqrt{\mathcal{R}}\right)\hat{b}_{0}^{\dagger}\right]+\cosh(s_{1})e^{i\varphi_{2}}\hat{a}_{i}+\cosh(s_{2}){b}_{i}\right\rbrace \nonumber\tag{S15}
  \end{equation}	
 \begin{equation}\label{general output}
	\hat{j}=\frac{1}{\sqrt{2}}\left\lbrace \left[\left(\sinh(s_{1})\sqrt{\mathcal{R}}e^{i\varphi_{2}}+\sinh(s_{2})\sqrt{1-\mathcal{R}}\right)\hat{a}_{0}^{\dagger}+\left(\sinh(s_{1})\sqrt{1-\mathcal{R}}e^{i\varphi_{\varphi_{2}}}-\sinh(s_{2})\sqrt{\mathcal{R}}\right)\hat{b}_{0}^{\dagger}\right]+\cosh(s_{1})e^{i\varphi_{2}}\hat{a}_{i}-\cosh(s_{2}){b}_{i}\right\rbrace \nonumber\tag{S16}
\end{equation}

\begin{figure}[h]
	\centering
	\includegraphics[width=0.9 \textwidth]{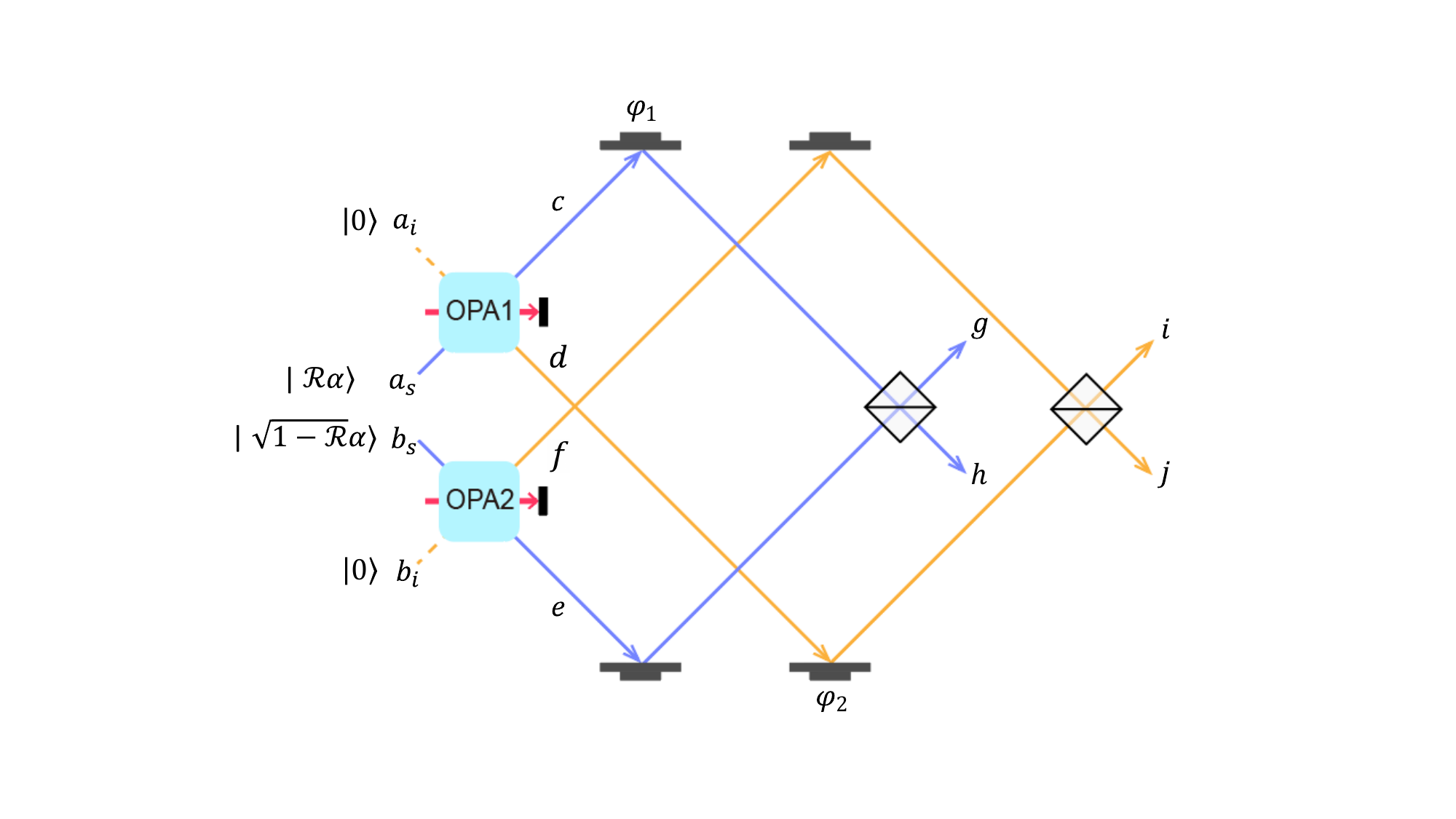} 
	\caption{\textbf{Schematic of the Quantum Correlated Interferometer.} }
	\label{fig.s3}
\end{figure}

The output differential currents in signal and idler modes become
\begin{equation}\label{hd output2}
\begin{split}
I_{1}&=\hat{g}^{\dagger}\hat{g}-\hat{h}^{\dagger}\hat{h}\nonumber
\\&=-2\cosh(s_{1})\cosh(s_{2})\sqrt{\mathcal{R}(1-\mathcal{R})}\cos(\varphi_{1})\alpha^{2}+\cosh(s_{1})\cosh(s_{2})\left[\mathcal{R}\hat{X}_{b_{0}}(\varphi_{1})-(1-\mathcal{R})\hat{X}_{b_{0}}(-\varphi_{1}) \right]\alpha\nonumber\\ &+\left[\cosh(s_{1})\sinh(s_{2})\sqrt{\mathcal{R}}\hat{X}_{bi}(-\varphi_{1})-\cosh(s_{2})\sinh(s_{1})\sqrt{1-\mathcal{R}}\hat{X}_{ai}( \varphi_{1})\right]\alpha \nonumber
\end{split}\tag{S17}
\end{equation}
\begin{equation}\label{hd output2}
\begin{split}
I_{2}&=\hat{i}^{\dagger}\hat{i}-\hat{j}^{\dagger}\hat{j}\nonumber\\&=-2\sinh(s_{1})\sinh(s_{2})\sqrt{\mathcal{R}(1-\mathcal{R})}\cos(\varphi_{2})\alpha^{2} +\sinh(s_{1})\sinh(s_{2})\left[\mathcal{R}\hat{X}_{b_{0}}(-\varphi_{2})-(1-\mathcal{R})\hat{X}_{b_{0}}(\varphi_{2}) \right]\alpha\nonumber\\& +\left[ \cosh(s_{2})\sinh(s_{1})\sqrt{\mathcal{R}}\hat{X}_{bi}(\varphi_{2})-\cosh(s_{1})\sinh(s_{2})\sqrt{1-\mathcal{R}}\hat{X}_{ai}(-\varphi_{2})\right]\alpha
\end{split}\tag{S18}
\end{equation}
Here $ \hat{X}_{\beta}(\gamma)=\hat{\beta}e^{-i\gamma}+\hat{\beta}^{\dagger}e^{i\gamma} $ with $ \beta\in\left\lbrace b_{0},a_{i},b_{i} \right\rbrace  $ are the quadrature of the input states, and $ \gamma $ is the phase of the quadrature. Then we can get the sum of the differential currents

\begin{equation}
\begin{split}
	I&= I_{1}+ I_{2} \nonumber\\&=-2\sqrt{\mathcal{R}(1-\mathcal{R})}\left[\cosh(s_{1})\cosh(s_{2})\cos(\varphi_{1})+\sinh(s_{1})\sinh(s_{2})\cos(\varphi_{2})\right]\alpha^{2}\nonumber\\& +\mathcal{R}\left[\cosh(s_{1})\cosh(s_{2})\hat{X}_{b_{0}}(\varphi_{1})+\sinh(s_{1})\sinh(s_{2})\hat{X}_{b_{0}}(-\varphi_{2}) \right]\alpha\nonumber\\& -(1-\mathcal{R})\left[\cosh(s_{1})\cosh(s_{2})\hat{X}_{b_{0}}(-\varphi_{1})+\sinh(s_{1})\sinh(s_{2})\hat{X}_{b_{0}}(\varphi_{2}) \right]\alpha\nonumber\\ &+\sqrt{\mathcal{R}}\left[\cosh(s_{1})\sinh(s_{2})\hat{X}_{b_{i}}(-\varphi_{1})+\cosh(s_{2})\sinh(s_{1})\hat{X}_{b_{i}}(\varphi_{2}) \right]\alpha\nonumber\\& -\sqrt{1-\mathcal{R}}\left[\cosh(s_{1})\sinh(s_{2})\hat{X}_{a_{i}}(-\varphi_{2})+\cosh(s_{2})\sinh(s_{1})\hat{X}_{a_{i}}(\varphi_{1}) \right]\alpha  
 \end{split}\tag{S19}
\end{equation}
The optimal SNR is achieved when $ \varphi_{1}=\varphi_{10}+\delta\varphi_{1} $ and $ \varphi_{2}=\varphi_{20}+\delta\varphi_{2} $ with the operating point $ \varphi_{10}=\pi/2 $, $ \varphi_{20}=\pi/2 $, $ \delta\varphi_{1}\rightarrow0 $ and $ \delta\varphi_{2}\rightarrow0 $, allowing us to acquire the maximum signal and minimal noise simultaneously
\begin{equation}
\begin{split}
\left\langle\Delta I\right\rangle^{2}&=(\left\langle I\right\rangle -\left\langle I\right\rangle_{0} )^{2}\nonumber\\&=\left\lbrace2\sqrt{\mathcal{R}(1-\mathcal{R})}\left[\cosh(s_{1})\cosh(s_{2})\delta\varphi_{1}+\sinh(s_{1})\sinh(s_{2})\delta\varphi_{2}\right]\alpha^{2}\right\rbrace^{2}
\end{split}\tag{S20}
\end{equation}
\begin{equation}
\begin{split}
	\left\langle \delta^{2}I\right\rangle_{0}=\left\lbrace\left[\cosh(s_{1})\cosh(s_{2})-\sinh(s_{1})\sinh(s_{2})\right]+\left[\cosh(s_{2})\sinh(s_{1})-\cosh(s_{1})\sinh(s_{2})\right]\right\rbrace\alpha^{2}
\end{split}\tag{S21}
\end{equation}
Here $ \left\langle I\right\rangle _{0} $ represents the expected output with $ \varphi_{1}=\varphi_{10} $ and $ \varphi_{2}=\varphi_{20} $, denoting an undisturbed interferometer. $ \left\langle I\right\rangle  $ denotes the expectation with $ \varphi_{1}=\varphi_{10}+\delta\varphi_{1} $ and $ \varphi_{2}=\varphi_{10}+\delta\varphi_{2} $, in the presence of weak signals causing slight phase excursion in the interferometer. Finally, we get the SNR for the PQCI
\begin{equation}
\zeta_{PQCI}=\frac{\left\lbrace2\sqrt{\mathcal{R}(1-\mathcal{R})}\left[\cosh(s_{1})\cosh(s_{2})\delta\varphi_{1}+\sinh(s_{1})\sinh(s_{2})\delta\varphi_{2}\right]\alpha^{2}\right\rbrace^{2}}{\left\lbrace\left[\cosh(s_{1})\cosh(s_{2})-\sinh(s_{1})\sinh(s_{2})\right]+\left[\cosh(s_{2})\sinh(s_{1})-\cosh(s_{1})\sinh(s_{2})\right]\right\rbrace\alpha^{2}}\tag{S22}
\end{equation}

The SNR reaches maximum when $ \mathcal{R}=1/2$, which is equal to the SNR of dual-beam SU(1,1) interferometer (SUI) [34]. In this case, we call it QTI. When the signals are induced with common-mode phase $ \delta\varphi_{1}=\delta\varphi_{2}\rightarrow\delta\varphi_{sig}  $ and $s_{1}=s_{2}=s$, we get the minimum sensitivity in phase measurement
\begin{equation}
\delta\varphi_{m}=\frac{1}{\sqrt{\cosh(2s)I_{ps}}}\tag{S23}
\end{equation}

Here $ I_{ps}=\left[\cosh^{2}(s)+\sinh^{2}(s)\right]\alpha^{2}=\cosh(2s)\alpha^{2} $ is the power of the phase sensing field. When $\alpha\rightarrow1$, we find the sensitivity approach the Heisenberg limit

\begin{equation}
\delta\varphi_{HL}=\frac{1}{I_{ps}}\tag{S24}
\end{equation}
The difference between tSUI and QTI primarily arises from different detection strategies. Here, we directly compare the results from measuring entangled probe beams with homodyne detection (HD) and entangled detection (ED). Unlike the description in the main text, where the total optical intensity is kept constant, here we fix the intensity of the entangled probe beam while varying the intensity of the reference beam in the two detection schemes. The power ratio $\mathcal{R}$, which represents the power of entangled probe beam relative to the total interference power. For EDs, $\mathcal{R}=|\alpha_{1}|^{2}/(|\alpha_{1}|^{2}+|\alpha_{2}|^{2})$. For HD, $\mathcal{R}=|\cosh(2s)\beta|^{2}/(\cosh(2s)|\beta|^{2}+|\alpha_{L}|^{2})$. The signals for both HD and ED exhibit the same dependence on $\mathcal{R}$. However, in HD, where coherent light is used as the reference, the quantum advantage gradually diminishes with the increase of $\mathcal{R}$. In contrast, in EDs, since the reference is also entangled, the noise remains independent of $\mathcal{R}$. 

\section{Quantum Twin interferometer with losses}
In practical experiments, losses are unavoidable. Here, we discuss the losses caused by optical path and mode mismatch in interferometry. In such cases, the input-output relation changes to
\begin{equation}\label{Input-output relation}
\hat{c}=\sqrt{1-\kappa_{s}}\left[\cosh(s)\hat{a}_{s}+\sinh(s)\hat{a}_{i}^{\dagger}\right]+\sqrt{\kappa_{s}}\hat{L}_{sv}\tag{S25}\\
\end{equation}
\begin{equation}\label{Input-output relation}
\hat{d}=\sqrt{1-\kappa_{i}}\left[\cosh(s)\hat{b}_{i}+\sinh(s)\hat{b}_{s}^{\dagger}\right]+\sqrt{\kappa_{i}}\hat{L}_{iv}\tag{S26}\\
\end{equation}
\begin{equation}\label{Input-output relation}
\hat{e}=\sqrt{1-\kappa_{s}}\left[\cosh(s)\hat{b}_{s}+\sinh(s)\hat{b}_{i}^{\dagger}\right]+\sqrt{\kappa_{s}}\hat{L}_{sv}\tag{S27}\\
\end{equation}
\begin{equation}\label{Input-output relation}
\hat{f}=\sqrt{1-\kappa_{i}}\left[\cosh(s)\hat{a}_{i}+\sinh(s)\hat{a}_{s}^{\dagger}\right]+\sqrt{\kappa_{i}}\hat{L}_{iv}\tag{S28}\\
\end{equation}
\begin{equation}\label{Input-output relation}
\hat{g}=\frac{1}{\sqrt{2}}\sqrt{1-\sigma_{s}}(e^{i\varphi_{1}}\hat{c}+\hat{e})+\sqrt{\sigma_{s}}\hat{L}_{st}\tag{S29}\\
\end{equation}
\begin{equation}\label{Input-output relation}
\hat{h}=\frac{1}{\sqrt{2}}\sqrt{1-\sigma_{s}}(e^{i\varphi_{1}}\hat{c}-\hat{e})+\sqrt{\sigma_{s}}\hat{L}_{st}\tag{S30}\\
\end{equation}
\begin{equation}\label{Input-output relation}
\hat{i}=\frac{1}{\sqrt{2}}\sqrt{1-\sigma_{i}}(e^{i\varphi_{2}}\hat{f}+\hat{d})+\sqrt{\sigma_{i}}\hat{L}_{it}\tag{S31}\\
\end{equation}
\begin{equation}\label{Input-output relation}
\hat{j}=\frac{1}{\sqrt{2}}\sqrt{1-\sigma_{i}}(e^{i\varphi_{2}}\hat{f}-\hat{d})+\sqrt{\sigma_{i}}\hat{L}_{it}\tag{S32}
\end{equation}
Here $\kappa_{s}$ and $\kappa_{i}$ represent losses from the optical path, while $\sigma_{s}$ and $\sigma_{i}$ represent losses from mode mismatch as illustrated in Fig.~\ref{fig_s2}. $\hat{L}_{t}$ with $t\in\left\lbrace sv,iv\right\rbrace$ are the vacuum noise induced from the optical path loss and $t\in\left\lbrace st,it\right\rbrace$ are the thermal noise induced from the mode mismatch. The output modes are simplified to
\begin{equation}
\begin{split}
	\hat{g}&=\frac{\sqrt{(1-\kappa_{s})(1-\sigma_{s})}}{\sqrt{2}}\left\lbrace\cosh(s)\left[\left(\sqrt{\mathcal{R}}e^{i\varphi_{1}} -\sqrt{1-\mathcal{R}}\right)\hat{a}_{0}+\left(\sqrt{1-\mathcal{R}}e^{i\varphi_{1}}+\sqrt{\mathcal{R}}\right)\hat{b}_{0}\right]+\sinh(s)\left(e^{i\varphi_{1}}\hat{a}_{i}^{\dagger}+{b}_{i}^{\dagger}\right)\right\rbrace \nonumber\\
	 &+\frac{\sqrt{\kappa_{s}(1-\sigma_{s})}}{\sqrt{2}}(1+e^{i\varphi_{1}})\hat{L}_{sv}+\sqrt{\sigma_{s}}\hat{L}_{st}
  \end{split}\tag{S33}
  \end{equation}
  \begin{equation}
\begin{split}
	\hat{h}&=\frac{\sqrt{(1-\kappa_{s})(1-\sigma_{s})}}{\sqrt{2}}\left\lbrace\cosh(s)\left[\left(\sqrt{\mathcal{R}}e^{i\varphi_{1}} -\sqrt{1-\mathcal{R}}\right)\hat{a}_{0}+\left(\sqrt{1-\mathcal{R}}e^{i\varphi_{1}}+\sqrt{\mathcal{R}}\right)\hat{b}_{0}\right]+\sinh(s)\left(e^{i\varphi_{1}}\hat{a}_{i}^{\dagger}+{b}_{i}^{\dagger}\right)\right\rbrace \nonumber\\
	&  +\frac{\sqrt{\kappa_{s}(1-\sigma_{s})}}{\sqrt{2}}(e^{i\varphi_{1}}-1)\hat{L}_{sv}+\sqrt{\sigma_{s}}\hat{L}_{st}
 \end{split}\tag{S34}
  \end{equation}
  \begin{equation}
\begin{split}
	\hat{i}&=\frac{\sqrt{(1-\kappa_{i})(1-\sigma_{i})}}{\sqrt{2}}\left\lbrace \sinh(s)\left[\left(\sqrt{\mathcal{R}}e^{i\varphi_{2}}-\sqrt{1-\mathcal{R}}\right)\hat{a}_{0}^{\dagger}+\left(\sqrt{1-\mathcal{R}}e^{i\varphi_{\varphi_{2}}}+\sqrt{\mathcal{R}}\right)\hat{b}_{0}^{\dagger}\right]+\cosh(s)\left(e^{i\varphi_{2}}\hat{a}_{i}+{b}_{i}\right)\right\rbrace \nonumber\\
	&  +\frac{\sqrt{\kappa_{i}(1-\sigma_{i})}}{\sqrt{2}}(1+e^{i\varphi_{2}})\hat{L}_{iv}+\sqrt{\sigma_{i}}\hat{L}_{it}\\
 \end{split}\tag{S35}
  \end{equation}
  \begin{equation}
\begin{split}
	\hat{j}&=\frac{\sqrt{(1-\kappa_{i})(1-\sigma_{i})}}{\sqrt{2}}\left\lbrace \sinh(s)\left[\left(\sqrt{\mathcal{R}}e^{i\varphi_{2}}+\sqrt{1-\mathcal{R}}\right)\hat{a}_{0}^{\dagger}+\left(\sqrt{1-\mathcal{R}}e^{i\varphi_{2}}-\sqrt{\mathcal{R}}\right)\hat{b}_{0}^{\dagger}\right]+\cosh(s)\left(e^{i\varphi_{2}}\hat{a}_{i}-{b}_{i}\right)\right\rbrace\nonumber\\
	&  +\frac{\sqrt{\kappa_{i}(1-\sigma_{i})}}{\sqrt{2}}(e^{i\varphi_{2}}-1)\hat{L}_{iv}+\sqrt{\sigma_{i}}\hat{L}_{it}
\end{split}\tag{S36}
  \end{equation}
\begin{figure}[h]
	\centering
	\includegraphics[width=0.9 \textwidth]{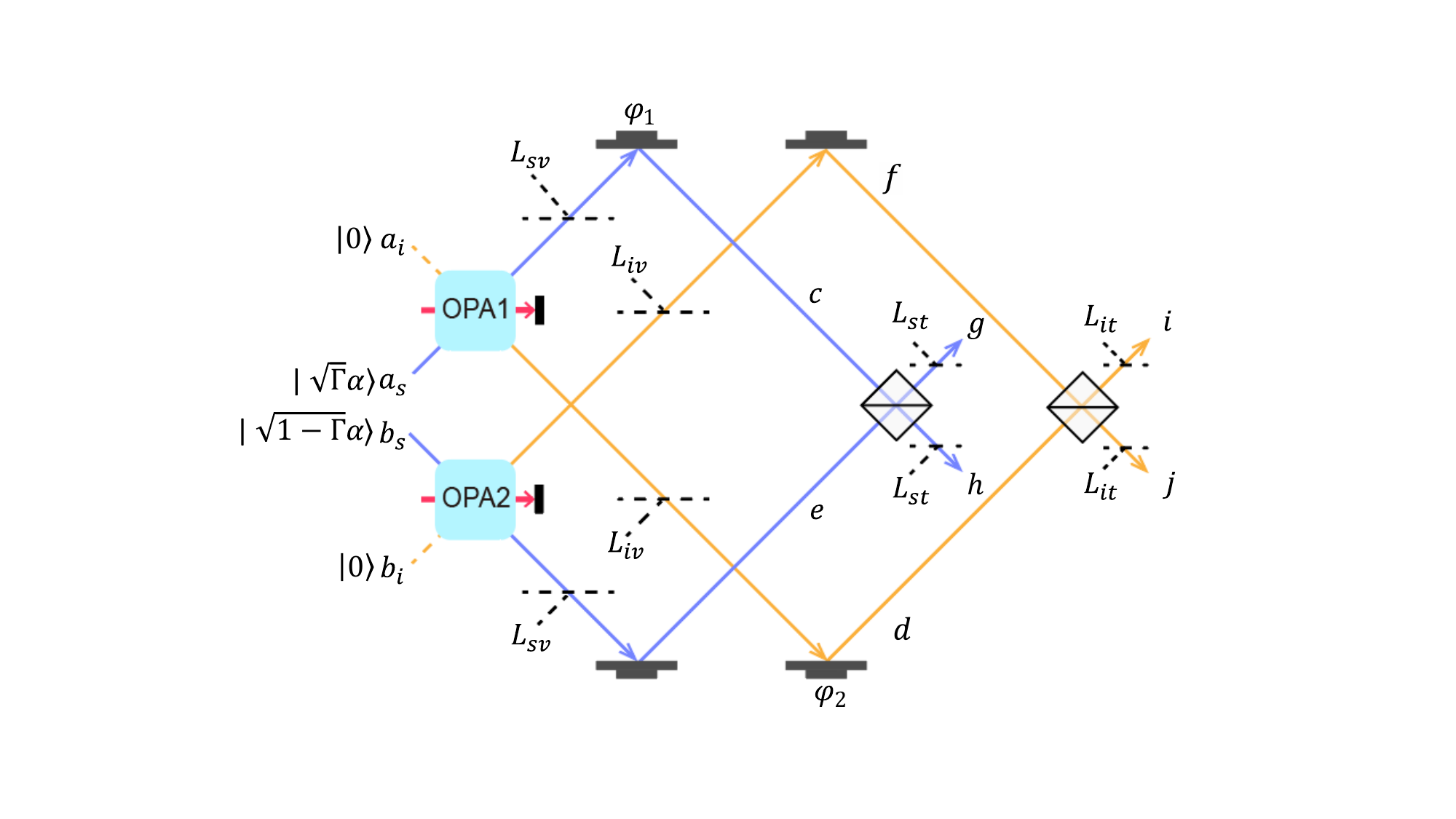} 
	\caption{\textbf{Twin Interferometer when losses are considered. }}
	\label{fig_s2}
\end{figure}

The output differential current in signal and idler modes become
 \begin{equation}
\begin{split}\label{hd output2}
\hat{I}_{1}&=\hat{g}^{\dagger}\hat{g}-\hat{h}^{\dagger}\hat{h}\nonumber\\&=(1-\kappa_{s})(1-\sigma_{s})\left\lbrace -2\cosh^{2}(s)\sqrt{\mathcal{R}(1-\mathcal{R})}\cos(\varphi_{1})\alpha^{2}+\cosh^{2}(s)\left[\mathcal{R}\hat{X}_{b_{0}}(\varphi_{1})-(1-\mathcal{R})\hat{X}_{b_{0}}(-\varphi_{1}) \right]\alpha\right.\nonumber\\ &\left.+\cosh(s)\sinh(s)\left[\sqrt{\mathcal{R}}\hat{X}_{bi}(-\varphi_{1})-\sqrt{1-\mathcal{R}}\hat{X}_{ai}( \varphi_{1})\right]\alpha\right\rbrace-(1-\sigma_{s})\sqrt{(1-\mathcal{R})\kappa_{s}(1-\kappa_{s})}\cosh(s)\hat{X}_{Lsv}(-\varphi_{1})\alpha \nonumber\\& -\sqrt{2R\sigma_{s}(1-\kappa_{s})(1-\sigma_{s})}\cosh(s)\hat{X}_{Lst}(0)\alpha
 \end{split}\tag{S37}
  \end{equation}
  \begin{equation}
\begin{split}\label{hd output2}
\hat{I}_{2}&=\hat{i}^{\dagger}\hat{i}-\hat{j}^{\dagger}\hat{j}\nonumber\\&=(1-\kappa_{i})(1-\sigma_{i})\left\lbrace -2\sinh^{2}(s)\sqrt{\mathcal{R}(1-\mathcal{R})}\cos(\varphi_{2})\alpha^{2} +\sinh^{2}(s)\left[\mathcal{R}\hat{X}_{b_{0}}(-\varphi_{2})-(1-\mathcal{R})\hat{X}_{b_{0}}(\varphi_{2}) \right]\alpha\right.\nonumber\\&\left.+\cosh(s)\sinh(s)\left[ \sqrt{\mathcal{R}}\hat{X}_{bi}(\varphi_{2})-\sqrt{1-\mathcal{R}}\hat{X}_{ai}(-\varphi_{2})\right]\alpha\right\rbrace-(1-\sigma_{i})\sqrt{(1-\mathcal{R})\kappa_{i}(1-\kappa_{i})}\sinh(s)\hat{X}_{Liv}(-\varphi_{2})\alpha \nonumber\\& -\sqrt{2R\sigma_{i}(1-\kappa_{i})(1-\sigma_{i})}\sinh(s)\hat{X}_{Lit}(0)\alpha
\end{split}\tag{S38}
  \end{equation}

The optimal SNR can be acquired when $\mathcal{R}=1/2 $, $ \varphi_{1}\rightarrow\varphi_{10}+\delta\varphi_{1} $ and $\varphi_{2}\rightarrow\varphi_{20}+\delta\varphi_{2} $ with the operating point $ \varphi_{10}=\varphi_{20}=\pi/2 $,  and weak signal $ \delta\varphi_{1}\rightarrow0 $, $ \delta\varphi_{2}\rightarrow0 $. There might be uneven losses among the interferometers in signal and idler modes due to different optical paths, leading to varying intensity and fluctuations between the currents $I_{1}$ and $I_{2}$. 
To address this issure, an electrical attenuation with factor $\eta=[\sinh(s)(1-\sigma_{i})(1-\kappa_{i})]/[\cosh(s)(1-\sigma_{s})(1-\kappa_{s})]$ is introduced to equalize the losses, yielding the subsequent output
\begin{equation}
\begin{split}\label{joint quadrature}
	\hat{I}&= \eta \hat{I}_{1}+ \hat{I}_{2}\nonumber\\&=(1-\kappa_{i})(1-\sigma_{i})\sinh(s)\alpha\left\lbrace\left[\cosh(s)\delta\varphi_{1}+\sinh(s)\delta\varphi_{2}\right]\alpha+e^{-s}\hat{X}_{b_{0}}(\pi/2)\right\rbrace \nonumber\\& -\frac{(1-\sigma_{i})(1-\kappa_{i})}{(1-\sigma_{s})(1-\kappa_{s})} \sinh(s)\alpha\left[ (1-\sigma_{s})\sqrt{\frac{\kappa_{s}(1-\kappa_{s})}{2}}\hat{X}_{Lsv}(-\pi/2)+\sqrt{\sigma_{s}(1-\kappa_{s})(1-\sigma_{s})}\hat{X}_{Lst}(0)\right]\nonumber\\& -\sinh(s)\alpha \left[ (1-\sigma_{i})\sqrt{\frac{\kappa_{i}(1-\kappa_{i})}{2}}\hat{X}_{Liv}(-\pi/2)+\sqrt{\sigma_{i}(1-\kappa_{i})(1-\sigma_{i})}\hat{X}_{Lit}(0) \right]
\end{split}\tag{S39}
  \end{equation}
We get the signal
\begin{equation}\label{joint quadrature}
	\langle\Delta I\rangle^{2}= \sinh^{2}(s)\alpha^{2}\left\lbrace(1-\kappa_{i})(1-\sigma_{i})\left[\cosh(s)\delta\varphi_{1}+\sinh(s)\delta\varphi_{2}\right]\alpha\right\rbrace^{2}\tag{S40}
\end{equation}
Here, the modes $\hat{L}_{sv}$ and $\hat{L}_{iv}$ represent vacuum states, while the modes $\hat{L}_{st}$ and $\hat{L}_{it}$ correspond to thermal states arising from mode-mismatch in interferometry. As a result, the noise levels are $\left\langle \delta^{2}\hat{X}_{Lsv}\right\rangle=1$ and $\left\langle \delta^{2}\hat{X}_{Liv}\right\rangle=1$. Additionally, $\left\langle \delta^{2}\hat{X}_{Lst}\right\rangle=e^{2s}$ and $\left\langle \delta^{2}\hat{X}_{Lit}\right\rangle=e^{2s}$, with their fluctuations depending on the gain of the PA processes. This leads to the determination of the noise performance of QTI at the output
\begin{equation}
\begin{split}
\left\langle \delta^{2}\hat{I}\right\rangle&=\sinh^{2}(s)\alpha^{2}\left\lbrace\left[e^{-s}(1-\kappa_{i})(1-\sigma_{i})\right]^{2}+\left[ \frac{(1-\sigma_{i})(1-\kappa_{i})}{(1-\sigma_{s})(1-\kappa_{s})}\right]^{2}\left[ \frac{\kappa_{s}(1-\kappa_{s})(1-\sigma_{s})^{2}}{2}+e^{2s}\sigma_{s}(1-\kappa_{s})(1-\sigma_{s})\right]\right.\nonumber\\&\left.+\left[ \frac{\kappa_{i}(1-\kappa_{i})(1-\sigma_{i})^{2}}{2}+e^{2s}\sigma_{i}(1-\kappa_{i})(1-\sigma_{i})\right] \right\rbrace
\end{split}\tag{S41}
  \end{equation}
Finally, we get the SNR
\begin{equation}
\begin{split}\label{snrtil}
\zeta_{QTI}&=\frac{\left\lbrace(1-\kappa_{i})(1-\sigma_{i})\left[\cosh(s)\delta\varphi_{1}+\sinh(s)\delta\varphi_{2}\right]\alpha\right\rbrace^{2}}{M}
\end{split}\tag{S42}
  \end{equation}
$M=\left[e^{-s}(1-\kappa_{i})(1-\sigma_{i})\right]^{2}+\left[ (1-\sigma_{i})(1-\kappa_{i})/(1-\sigma_{s})(1-\kappa_{s})\right]^{2}\left[ \kappa_{s}(1-\kappa_{s})(1-\sigma_{s})^{2}/{2}+e^{2s}\sigma_{s}(1-\kappa_{s})(1-\sigma_{s})\right]+\left[ \kappa_{i}(1-\kappa_{i})(1-\sigma_{i})^{2}/{2}+e^{2s}\sigma_{i}(1-\kappa_{i})(1-\sigma_{i})\right] $. For QTI, the power of phase sensing light is $I_{QTI}=\cosh(2s)\alpha^{2}/2$. Our system can be easily converted to the conventional Mach-Zehnder interferometer (MZI) by setting $s=0$. For a fair comparison, the weak signal in MZI should satisfy $\delta\varphi_{mz}=\delta\varphi_{1}/2+\delta\varphi_{2}/2$, and from which we obtain the SNR for MZI
\begin{equation}
\label{snrmzl}
	\zeta_{MZI}=\frac{\left[2(1-\kappa_{s})(1-\sigma_{s})\delta\varphi_{mz}\alpha\right]^{2}}{\left[(1-\kappa_{s})(1-\sigma_{s})\right]^{2}+\left[ \frac{\kappa_{s}(1-\kappa_{s})(1-\sigma_{s})^{2}}{2}+\sigma_{s}(1-\kappa_{s})(1-\sigma_{s})\right] }\tag{S43}
\end{equation}
Here, the power of phase sensing light for MZI is $I_{MZI}=\alpha^{2}/2$. Eq.~\ref{snrtil} and Eq.~\ref{snrmzl} are used in fitting to the measured data, noting that all traces in Fig. 4 meet the condition $I_{QTI}=I_{MZI}$ for a fair comparison. We find that the theory aligns well with the experimental results. The optimal fitting parameters are $\kappa_{s}=0.2$, $\kappa_{i}=0.1$, $\sigma_{s}=0.03$, and $\sigma_{i}=0.02$.

\section{Energy level diagrams.}

Energy level diagram in the $D_{1}$ line of $^{85}$Rb for FWM process is given in the Fig.~\ref{Energy level}. Here $ \Delta$ is the one-photon detuning, and $ \delta$ is two-photon detuning. P$_{1}$ denotes the FWM pump field.

\begin{figure}[h]
	\centering
	\includegraphics[width=0.3 \textwidth]{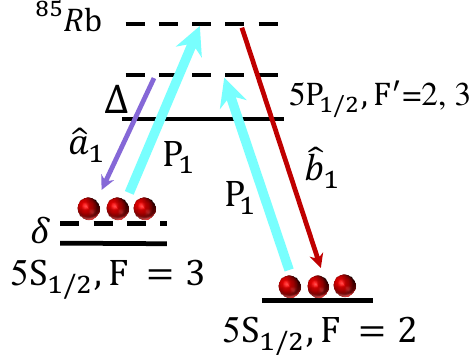} 
	\caption{\textbf{Energy level diagrams of $^{85}$Rb.} }
	\label{Energy level}
\end{figure}
\include{references}

\end{widetext}
\end{document}